\documentclass[fleqn,usenatbib]{mnras}

\usepackage{mathptmx}

\usepackage[T1]{fontenc}
\usepackage{ae,aecompl}

\usepackage{graphicx}	
\usepackage{amsmath}	
\usepackage{amssymb}	
\usepackage{lscape} 
\usepackage{url} 

\newcommand{\HI}{H\,{\sc{i}}} 

\newcommand{\Msun}{$\mathrm{M_{\odot}}$} 

\newcommand{\kms}{km\,s$^{-1}$}

\title[The Fornax cluster and beyond]{A spectroscopic census of the Fornax
  cluster and beyond: preparing for next generation surveys}

\author[N. Maddox et al.]{
  Natasha Maddox $^{1,2}$\thanks{E-mail: natasha.maddox@gmail.com}, 
Paolo Serra $^{3}$, Aku Venhola $^{4}$, Reynier Peletier $^{5}$,
\newauthor Ilani Loubser $^{6}$, Enrichetta Iodice $^{7}$
  \\
$^{1}$Faculty of Physics, Ludwig-Maximilians-Universit\"at,
Scheinerstr. 1, 81679 Munich, Germany\\
$^{2}$ASTRON, the Netherlands Institute for Radio Astronomy, Oude
Hoogeveensedijk 4, 7991 PD, Dwingeloo, The Netherlands\\
$^{3}$INAF - Osservatorio Astronomico di Cagliari, Via della Scienza 5,
I-09047 Selargius (CA), Italy\\
$^{4}$Astronomy Research Unit, University of Oulu, Pentti Kaiteran
katu 1, 90014 Oulu, Finland\\
$^{5}$Kapteyn Astronomical Institute, University of Groningen, PO Box
72, NL-9700 AB Groningen, The Netherlands\\
$^{6}$Centre for Space Research, North-West University, Potchefstroom, 2520,
South Africa\\
$^{7}$INAF-Astronomical Observatory of Capodimonte, via Moiariello 16,
Napoli, Italy\\ 
}

\date{Accepted XXX. Received YYY; in original form ZZZ}

\pubyear{2019}

\begin{document}
\label{firstpage}
\pagerange{\pageref{firstpage}--\pageref{lastpage}}
\maketitle

\begin{abstract}

The Fornax cluster is the nearest, large cluster in the southern
sky, and is currently experiencing active assembly of mass.
It is thus the target of a number of ongoing observing
campaigns at optical, near-infrared and radio wavelengths, using
state-of-the-art facilities in the Southern hemisphere.
Spectroscopic redshifts are essential not only for determining cluster
membership, but also kinematics within the cluster and identifying substructures.
We present a compilation of all available major spectroscopic campaigns
undertaken within the Fornax region, including new and previously
unpublished spectroscopy. This provides not only a comprehensive
census of Fornax cluster membership as a resource for the many ongoing
studies of this dynamic system, but also probes the large scale
structure in the background volume.

\end{abstract}



\section{Introduction}\label{sec:Introduction}

The observable properties of galaxies are known to be strongly
dependent on their local and large-scale environment
(e.g. \citealt{Dressler1980}; \citealt{Binggeli1990}; \citealt{Peng2010}).
Clusters, as the densest environment, thus provide an excellent
laboratory for investigating environmental effects felt by
galaxies, where we are able to witness the transformational processes such as
interactions and mergers, and stripping of the gaseous component of
galaxies. 

The Fornax cluster, located at a distance of 20.0 megaparsecs (Mpc; 
\citealt{Blakeslee2009}), is the second most massive cluster within 20
Mpc, and is the largest nearby cluster in the Southern
hemisphere. It is dominated by an overdensity centred on the brightest
cluster galaxy (BCG) NGC 1399, and has a lower-density substructure
surrounding the radio galaxy NGC 1316 (hereinafter referred to as 
Fornax A) located to the south-west. It's relative proximity 
means galaxy interactions, and processes of
accretion/removal of gas onto/from galaxies can be studied in great
detail, with observations extending back at least 60 years
(e.g. \citealt{Hodge1959}). Despite the long history of studies 
targeting the general galaxy population, along with specific
sub-populations, there is renewed interest in the Fornax cluster,
coinciding with the availability of new observing facilities in the south.

The Fornax cluster has recently been at the centre of several major
observing programmes over a wide range of wavelengths. This includes
deep optical imaging (\citealt{Iodice2016}; \citealt{Venhola2018}; 
\citealt{Eigenthaler2018}), optical integral field unit (IFU) 
observations of selected targets (\citealt{Scott2014};
\citealt{Mentz2016}; \citealt{Sarzi2018}), ALMA observations 
(\citealt{Zabel2019}; \citealt{Morokuma2019}), far-infrared Herschel
imaging (\citealt{Davies2013}) and, at radio wavelengths, a survey
carried out with the Australia Telescope Compact 
Array (ATCA; \citealt{LeeWaddell2018}).

The reason for this ongoing interest is the Fornax cluster's role in
our understanding of galaxy evolution in dense environments. 
Fornax is very different from other nearby clusters. Compared to the
well-known Virgo and Coma clusters, Fornax is about one and two orders
of magnitude smaller, respectively, in total halo mass
(\citealt{Drinkwater2001b} for Fornax; \citealt{Lokas2003} for Virgo), meaning the
density of the intra-cluster medium (ICM; \citealt{Paolillo2002} for
Fornax; \citealt{Fabricant1980} for Virgo) and the velocity dispersion
of its galaxy population are lower. Therefore, ram pressure stripping 
(a prime candidate for the removal of gas from galaxies and for the
quenching of star formation in dense environments), which scales as
the ICM density times the relative velocity squared, should be significantly
weaker in Fornax than in Coma and Virgo (i.e. an order of magnitude
lower compared to that in Virgo, see, for example
\citealt{Venhola2019}). Fornax also has a higher galaxy 
number density in its centre. Coupled with the lower velocity
dispersion, this creates an ideal environment for tidal interactions
to be effective.

The Fornax cluster is thus an environment in which the balance
between hydrodynamical and tidal effects is quite different compared
to other, better-studied clusters. It is therefore an important system
for extending the baseline of cluster properties over which we have a
detailed view of galaxy transformation in action. And it is indeed in
action, since Fornax continues to grow by accreting new galaxies and
galaxy groups \citep{Drinkwater2001b}. The best example of accretion
is the relatively gas-rich group centred on Fornax A, currently at the
edge of the cluster but moving towards it, possibly along a large
scale filament of the cosmic web (\citealt{Scharf2005};
\citealt{Venhola2019}). 

Hindering any study of galaxy clusters is the availability of
spectroscopic redshifts, fundamental for not only confirming cluster membership, but
also deriving kinematics within the cluster. A number of authors have
performed observations within the Fornax cluster region targeting
particular sub-populations of objects, or specific spatial regions,
resulting in a large number of, albeit heterogeneously selected, 
spectroscopic redshifts (listed and described in
Section~\ref{sec:spectra}). Taken individually, no one survey gives
complete coverage of the cluster, but when combined, the projects 
result in a comprehensive census of the Fornax cluster galaxy
population, the foreground Galactic stars, as well as the background
galaxies in the volume behind the cluster.

The new surveys currently underway provide motivation to compile the
spectroscopy into one useful resource. Here we describe some of the
recent large programmes either already underway or about to begin,
which will directly benefit from such a compilation.

\subsection{The Fornax Deep Survey}\label{subsec:FDS}

The Fornax Deep Survey (FDS; \citealt{Iodice2016}; 
\citealt{Venhola2018}) is one of the new 
imaging surveys exploring the Fornax cluster. Using the VLT Survey
Telescope (VST) at the Paranal observatory of 
the European Southern Observatory (ESO), the FDS is part of Guaranteed
Time Observation surveys FOCUS (PI R Peletier) and VEGAS (PI E
Iodice). The 2.6-m wide-field VST, coupled with the 1 square 
degree field of view OmegaCam \citep{Kuijken2011} provides imaging
in the Sloan Digital Sky Survey (SDSS) $u, g, r$ and $i$ bands, with a
pixel scale of 0.21 arcsec pix$^{-2}$. 

The imaging survey was completed at the end of 2017, 
and resulted in images with surface brightness
sensitivity extending to 27--28 mag arcsec$^{-2}$ in the four optical bands. The
area covered spans 26 deg$^{2}$, centred on NGC 1399, and extending to
include Fornax A. The key goals of the FDS include investigating 
the ultra-diffuse galaxies (UDGs; \citealt{Venhola2017}) and 
the dwarf galaxy population of the cluster
(e.g. \citealt{Venhola2018}), a census of the globular cluster
population (\citealt{DAbrusco2016}; \citealt{Cantiello2018}), 
the low surface brightness outskirts of galaxies
(e.g. \citealt{Iodice2019}), along with the intra-cluster light, to
understand the formation history of galaxies within the cluster environment. 

\subsection{The MeerKAT Fornax Survey}\label{subsec:MFS}

The relatively low cluster cluster mass, resulting in low velocity
dispersion, along with the high density of galaxies, means the Fornax
cluster is an interesting laboratory within which to observe the
behaviour of the neutral hydrogen (\HI) component of galaxies, which
is a key tracer of processes that drive the evolution of galaxies in
dense environments. 

The MeerKAT Fornax Survey (MFS; \citealt{Serra2016}) will utilise the
newly-commissioned MeerKAT radio telescope \citep{Jonas2016} to
observe \HI\ and radio synchrotron emission of galaxies and of the
intra-cluster medium in the Fornax cluster. Observations spanning 12
deg$^2$ will cover the central NGC 1399 
region, extending towards the Fornax A subgroup, as seen in
Fig.~\ref{fig:footprint}. The supreme sensitivity of MeerKAT will
enable the low column density \HI\ extending beyond the stellar disks
to be probed, including streams of \HI\ removed from galaxies via
ram-pressure stripping. With a predicted \HI\ mass detection limit of
$\sim 5\times 10^5$\,\Msun, low-mass dwarf galaxies along with
\HI\,-poor galaxies will all be detected. The relative proximity of
the cluster results in 1 kiloparsec (kpc) spatial resolution, with
down to 1\,\kms\ spectral resolution. The MFS, coupled with the FDS
imaging, will thus be an ideal set of observations for understanding
the transformational processes underway within dense environments.

\subsection{MIGHTEE-\HI}\label{subsec:MIGHTEE-HI}

Due to the large bandwidth of the MeerKAT L-band receiver, the \HI\
spectral line is observable over the redshift range
$0<z_{\mathrm{HI}}<0.58$. The Fornax cluster occupies a very small slice of
this volume. In order to maximise the scientific return from the
MFS observations, the volume \textit{behind} the Fornax cluster
will be analysed as part of another MeerKAT large survey, the MeerKAT
International Giga-Hertz Tiered Extragalactic Exploration (MIGHTEE;
\citealt{Jarvis2016}). MIGHTEE is a commensal radio continuum and
spectral line survey, with the working group responsible for the \HI\
emission data designated as MIGHTEE-\HI, and an additional working group
focusing on \HI\ in absorption.

The depth of the Fornax observations is well-matched to 
the MIGHTEE observations, which results in predicted sensitivity
to direct detections of the \HI\ line to $z\sim 0.4$ (see Figure 2
from \citealt{Maddox2016}). Having a spectroscopic census of the
volume behind the Fornax cluster is of great use for the MIGHTEE \HI\
emission and absorption science, particularly at the higher redshift
end of the volume, where statistical techniques such as stacking will be used. 
\\
Here, we present new spectroscopic observations which substantially
add to the coverage of the FDS and MFS area, focusing on the
extragalactic population. We then include these new spectra in a 
compilation of 12 separate spectroscopic surveys of the
Fornax cluster region, including both published and unpublished
data. This compilation is not only the most comprehensive census
of Fornax cluster objects, but also includes redshifts for several
thousands of objects behind the Fornax volume, providing essential
data for the MIGHTEE survey.

The structure of the paper is as follows: in Section~\ref{sec:spectra} we
detail each of the works,
both published and unpublished, that contribute to our final
spectroscopic catalogue. Section~\ref{sec:FCandbeyond} investigates
the velocity structure of the Fornax cluster, and the various
structures found in the background volume. A discussion and summary
are given in Section~\ref{sec:discussion}. 
Concordance cosmology with $H_{0} = 70$ km s$^{-1}$ Mpc$^{-1}$ (thus $h\equiv
H_{0}$/[100 km s$^{-1}$ Mpc$^{-1}$]$=0.7$), $\Omega_{m} = 0.3$,
$\Omega_{\Lambda} = 0.7$ is assumed, and AB magnitudes are used throughout
unless otherwise stated.

\section{Input Spectroscopy}\label{sec:spectra}

The aim of the current work is to collect all available redshifts
within the Fornax cluster sky area, in order to determine cluster
membership as well as background volume large scale structure. 
This will serve as a useful resource not only for studies targeting
the cluster itself, but also work focused on the volume behind the
cluster. A major contribution to the compilation comes from newly acquired 
spectroscopy, which provides nearly one quarter of the redshifts of
extragalactic objects in the catalogue.

We have searched the literature for spectroscopic campaigns either
specifically targeting the Fornax cluster region, or large-area
surveys that include the Fornax cluster. A short description of each
source of spectroscopy is given below, along with the relevant
references for further, more detailed information. Observations
targeting single objects within the cluster, such as the Fornax 3D
project (\citealt{Sarzi2018}) are not included, as they do not
contribute to the redshift census.

Many of the spectroscopic campaigns collated here focused on the compact stellar
system (CSS) populations within the cluster, primarily associated with
the BCG NGC 1399. Therefore, there is a
large number of spectroscopically confirmed CSS objects.
As noted in several of the following works, there is some ambiguity
regarding the nomenclature for the various CSS populations 
found within the Fornax cluster. The populations within the
CSS umbrella term include globular clusters (GCs) bound to a host
galaxy, GCs not bound to a host galaxy, dwarf ellipticals (dEs) and
ultra-compact dwarf galaxies (UCDs). While UCDs tend to be more
massive, larger and brighter than GCs, deep observations show they
overlap in luminosity but not dynamically \citep{Gregg2009}. The
characteristics of the photometry may affect the 
morphological classification, with the input imaging ranging from 
digitised sky plates through new survey telescopes. Therefore, we use
the nomenclature native to each survey without further comment. While
GCs might not be considered galaxies unto themselves, we include
compilations of CSS, including GCs, for completeness.

\subsection*{This work}

\begin{figure}
\includegraphics[width=\columnwidth]{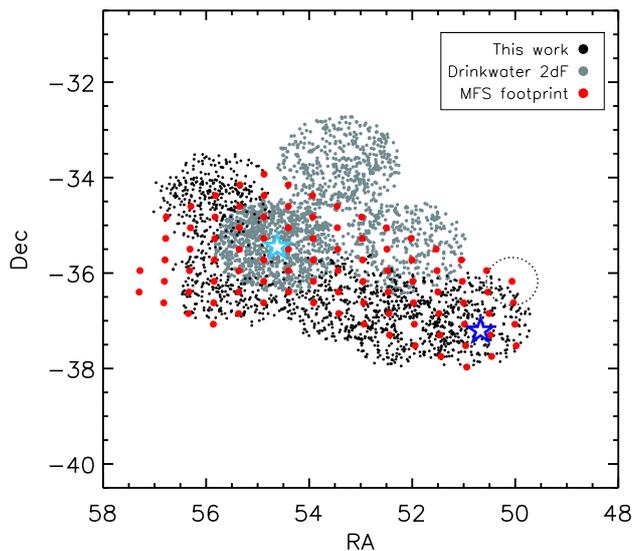}
\caption{Footprints of the 2dF pointings from this work (black points)
and Drinkwater 2dF (grey points), along with the outline of the area
to be covered by the MFS (red points). The red points denote the
centres of each MeerKAT pointing, each of which are $\sim$1 degree in
diameter, as indicated by the dotted circle, and are spaced by just
under 0.5 degrees from each other to ensure Nyquist sampling of the
MeerKAT primary beam. NGC 1399 is shown as the light blue star, while
the dark blue star indicates the location of Fornax A.}
\label{fig:footprint}
\end{figure}

While most observing campaigns in this area of the sky understandably focus
on cluster objects, the MIGHTEE-\HI\ survey is specifically surveying
the volume behind Fornax. Therefore, a census of background galaxies
to moderate redshifts is one of the aims of these observations. The
observations with the most comprehensive coverage of the Fornax
background volume is from \citet{Drinkwater2000a}, which covers $\sim$9
square degrees. We designed our observations to be complementary to
these, both in coverage and in depth. The footprints of the two
surveys are shown in Fig.~\ref{fig:footprint}. 

The input photometry is derived from the FDS (\citealt{Iodice2016}), with
catalogues of resolved objects provided by \citet{Venhola2018}. We impose a
magnitude limit of $g\le 19.5$ to be generally consistent with the
observations with the same instrument by \citet{Drinkwater2000a}. Employing the
catalogue of extended objects means there were very few Galactic stars
in our input catalogue.

We obtained three nights of observations in mixed conditions
with the AAOmega multi-object spectrograph \citep{Saunders2004}
and the Two-degree Field (2dF; \citealt{Lewis2002}) fibre positioner 
on the 3.9-m Anglo-Australian Telescope (AAT) 24-26 October 2017. We
employed the 580V and 385R gratings in the blue and red arms,
respectively, with central dispersion 1.03\AA\, pix$^{-1}$ and 1.56\AA\,
pix$^{-1}$, which results in velocity resolution of $\sim$ 220\,\kms. The
AAOmega spectrograph provides continuous wavelength 
coverage from 3800--8800\AA. Five 2dF fields were
observed, covering the MFS footprint while avoiding overlap with the
existing observations. The spectra were reduced using
2dfdr\footnote{The 2dfdr software is available from 
  \url{https://www.aao.gov.au/science/software/2dfdr}}
(\citealt{Croom2004}; \citealt{AAO2015}) and 
classified with the Manual and Automatic Redshifting Software MARZ
(\citealt{Hinton2016}). After visually inspecting the output of MARZ,
and altering the classification when necessary, we successfully obtained
redshifts for 1553 objects, 1549 of which are extragalactic, and 44
belong within the Fornax volume.

\subsection*{Drinkwater 2dF}

The Fornax Cluster Spectroscopic Survey (FCSS; \citealt{Drinkwater2000a}), was
also undertaken with the 2dF instrument on the AAT, before the
upgrade to the AAOmega spectrograph. Targets were
selected over $16.5<b_{J}<19.7$ from the UK Schmidt Telescope Sky
Survey plates, which have been digitised by the Automatic Plate
Measuring facility (APM) in Cambridge. Crucially, no morphological
selection was imposed on the input sample, which resulted in the
formalisation of a population of ultra-compact dwarf galaxies (UCDs) within the
Fornax cluster (\citealt{Drinkwater2000b}).

Four fields were planned within the FCSS, however only three were
observed, starting in 1997 and spanning multiple observing seasons. The
spectral coverage was 3600-8010\AA, with dispersion 
4.3\AA\ per pixel. We obtained the catalogue of objects for which a secure
redshift was determined from the authors (Drinkwater, private
communication). The catalogue contains both published and unpublished
redshifts. The lack of morphological selection results in a sample of
objects dominated by Galactic stars, with 4102 of 12071 objects
confirmed to be extragalactic, and 119 of these within the Fornax
volume. The location of the three Drinkwater 
2dF fields are also shown in Fig.~\ref{fig:footprint}. 

\subsection*{Firth 2dF}

\citet{Firth2008} targeted not only the core of the Fornax cluster,
centred on NGC 1399, but also the area around the neighbouring
overdensity at Fornax A, with the 2dF/AAOmega spectrograph on the
AAT in an observing campaign in December 2006. 
Targets were selected from APM-measured $b_{J}$ and $R$-band
photometry, choosing point sources with $b_{J}-R<1.7$, based on the
colours of previously identified compact galaxies within the Fornax
cluster. The magnitude limits for the area around Fornax A was 
$b_{J}<20.8$, and $b_{J}<21.8$ for the field centred on NGC 1399. 
The spectral coverage of the dual-beam AAOmega spectrograph was
3800--8900\AA. 

While the results from these observations have been published in
\citet{Firth2008}, the catalogue derived from the observations was
not. Therefore, we extracted the five nights of observations from the
AAT archive hosted by data
central\footnote{\url{datacentral.org.au}}. The data were reduced
using 2dfdr, in the same way as the spectra from this work, and
visually classified using MARZ.

For a number of observations, calibration files for one of the two
spectrograph arms were missing, so those observations were not reduced
or classified. No attempt to co-add multiple exposures of the same
target to increase the signal-to-noise ratio was made. As for the
Drinkwater 2dF observations, the selection of targets resulted in a
population of objects dominated by Galactic stars. Of 1906 objects
with secure redshifts, 352 are extragalactic, with 24 Fornax cluster members.

\subsection*{Gregg 2dF}

\citet{Gregg2009} also employed the 2dF spectrograph on the
AAT, in search of faint UCDs within the Fornax cluster centred on NGC 1399.
Targets were selected from the APM Catalogue $b_J$ and $r$ plates,
choosing unresolved objects with $b_{J}- r < 1.7$, over
$19.5<b_{J}<21.5$, extending to fainter magnitudes than the previously
known population.

Observations carried out in Oct 2003 and Nov 2004 resulted in
60 UCDs in the region of NGC 1399 (54 new discoveries), as well as 53
additional cluster members. The faint magnitude limit of the
observations result in UCDs overlapping in luminosity with NGC 1399
GCs, however the UCDs remain dynamically separate
\citep{Gregg2009}. Around half of the 
population is located in the intra-cluster volume, rather than
associated with a host galaxy, as GCs generally are. We add the 113
objects to our compilation.

\subsection*{Drinkwater FLAIR-II}

A spectroscopic campaign with the now-decommissioned FLAIR-II
multi-object spectrograph on the UK Schmidt Telescope at the Siding Spring
Observatory in Australia was undertaken over 1992--1997 by
\citet{Drinkwater2001a}. Restricted to $16.5<b_{J}<18$, these
observations are shallower, but cover a larger area than that probed
by the FCSS observations. 

Objects are again selected from the APM measurements of the $b_{J}$
survey plates, to be compact, but not stellar. Selection was based in
part on photometric classifications from the Fornax Cluster Catalog
(FCC, \citealt{Ferguson1989}). For observations
through 1996, the spectral coverage was 3670-7230\AA, with 13\AA\
resolution, and 5150-6680\AA, 5.3\AA\ resolution for observations in
1997. We extract the 108 confirmed Fornax cluster galaxies and 408
background galaxies from this work. 

\subsection*{Firth FLAMES}

Employing input imaging from the Cerro-Tololo Inter-American
Observatory (CTIO) Mosaic Imager with the 4-m Blanco Telescope in the
area around NGC 1399 (\citealt{Karick2006}); \citet{Firth2007}
selected point sources for 
observation with the Very Large Telescope Fibre Large Array Multi
Element Spectrograph (VLT-FLAMES; \citealt{Pasquini2002}). The point
sources were selected to be faint, extending to $r^{\prime} <22.75$, or
$b_{J}\la 23.25$, and $0.37<g^{\prime}-r^{\prime}<1.07$ and
$-0.06<r^{\prime}-i^{\prime}<0.64$ where $g^{\prime}$, $r^{\prime}$,
and $i^{\prime}$ are calibrated to the SDSS photometric system. Only
point sources were selected 
as the aim of the observations was to confirm CSS objects. 

Of 468 targets observed in November 2004, 98 Galactic stars and 57
Fornax objects were confirmed. The good resolution of the spectra,
with R$>$6000, resulted in recession velocity uncertainties of $cz_{err}\sim
15$\kms. The order of preference for consolidating the redshifts from
their Table 2 is VLT$>$Bergond$>$FCOS$>$2df.
In addition, 28 prominent galaxies near NGC 1399 with
redshifts from the literature are assembled. Based on brightness and
proximity to NGC 1399, the objects are separated into four categories,
including dwarf ellipticals, known cluster CSS, 
new cluster CSS, and other prominent galaxies, primarily early type
galaxies.

\subsection*{2MRS}

As the Fornax cluster is at low redshift, a number of the cluster galaxies
are bright and extended on the sky. Thus a number of cluster members
were observed within the Two Micron All Sky Survey (2MASS;
\citealt{Skrutskie2006}) Redshift Survey (2MRS; \citealt{Huchra2012}). 
Input photometry was supplied by the all-sky near-infrared imaging
from the 2MASS extended source catalogue (XSC; \citealt{Jarrett2000}. 
Spectrographs on multiple telescopes were employed, with spectral
coverage ranging from 3500--3700\AA\ at the blue end and 6400--7400\AA\
at the red end, sufficient to detect multiple spectral features of the
primarily low-redshift galaxies. New observations were supplemented by
existing spectroscopy from the literature.

Galaxies with $48<$RA$<59$ degrees and $-39<$Dec$<-33$ degrees were
extracted from both 
the main redshift catalogue, with $K_{s}\le 11.75$, and the additional
catalogue which extends to fainter magnitudes. 79 and 371 galaxies 
lie within the Fornax volume and background volume,
respectively.

\subsection*{Hilker 1999}

While the primary driver of the combined photometric and spectroscopic
campaign undertaken by \citet{Hilker1999a} (photometry) and
\citet{Hilker1999b} (spectroscopy) was to find dwarf galaxies near the
Fornax giant elliptical galaxies, the volume behind the Fornax cluster
was explored as well. 

Using the 2.5-m du Pont telescope at the Las Campanas Observatory,
images in Johnson $V$ and Cousins $I$ were obtained around several
Fornax spheroids, with a four-field mosaic centred on NGC
1399 (\citealt{Hilker1999a}). 125 targets to $V<20$ were selected for
spectroscopic follow-up with the multi-fibre spectrograph on the
du Pont telescope. Observations in December 1996 of 112 galaxies
resulted in confirmation of 7 Fornax members and 73 background
galaxies. Among the Fornax members were the first two UCD objects,
however they were not formally given this designation until
\citet{Drinkwater2000b}. These spectra confirmed the existence of a
background galaxy cluster at $z=0.11$, also clearly seen in our
compilation of spectra. Note that we use the `adopted' velocity from
their Table 2. 

\subsection*{Ferguson 1989}

\citet{Ferguson1989} undertook a seminal investigation of the Fornax
cluster. Image plates from the du Pont 2.5-m telescope, supplemented
with blue plates from the ESO UK
Science Research Council (SRC) survey taken on the UK Schmidt
Telescope provide the photometry over a $6^{\circ} \times 6^{\circ}$ region. 

Cluster members were identified by visual inspection, considering
their surface brightness, luminosity and morphology. This
classification resulted in a catalogue of 340 likely cluster members,
and 2338 background galaxies. The `FCC' designation still employed by
a number of authors to refer to specific galaxies originated within
this work.

Of the 340 galaxies presumed to belong to the Fornax cluster within
the original FCC catalogue, spectroscopy was compiled by
\citet{Ferguson1989} from various sources for 68 objects, confirming
their cluster membership.  Only these 68 objects are
included in our compilation. The B1950 coordinates of
the 68 objects have been converted to the J2000 epoch within the catalogue.

\subsection*{Bergond 2007}

Rather than focusing on a specific host galaxy, \citet{Bergond2007}
investigated the GC population between galaxies to better probe the
cluster dynamics.

Deep imaging of a strip centred on NGC 1399 with the WFI camera on the
ESO/MPG 2.2-m telescope provided the input imaging from which point
sources with $19.5\le V\le 22.2$ were selected as candidate
targets. Spectroscopy of more than 500 candidates was obtained with
the multi-object spectrograph FLAMES on the VLT. 149 GCs and 27 dwarf
galaxies within the Fornax cluster were confirmed, with 61 within the
intra-cluster medium.

\subsection*{Mieske 2004}

The Fornax Compact Object Survey (FCOS) undertaken by
\citet{Mieske2002} and \citet{Mieske2004} looked to further understand
the UCD galaxies within the Fornax cluster. Unresolved targets to
$V<21$ were
selected from imaging from the du Pont 2.5-m telescope and the CTIO
4-m telescope. Follow-up spectroscopy with the multi-slit mask Wide
Field CCD on the du Pont telescope in two campaigns, 30 Dec 2000 -- 1
Jan 2001, and 4--6 Dec 2002. The first employed a medium
resolution grism with spectral coverage 3500--6300\AA, giving 3\AA\
resolution, while the second used a lower resolution grism with wider
spectral coverage, 3600--7800\AA, for 6\AA\ resolution.

Combining the published results from the two sets of observations
results in 305 objects with confirmed velocities, including 65 Fornax
galaxies, of which 10 are UCDs, and 62 in the background volume. 

\subsection*{Schuberth 2010}

A comprehensive census of the GC population in the wide vicinity of
NGC 1399 has been undertaken by \citet{Schuberth2010}. Candidates for
spectroscopic followup were chosen from wide-field imaging from the
CTIO MOSAIC camera on the 4-m Blanco telescope. Multi-object
spectroscopy with the FORS2 instrument on the VLT (1--3 December 2002) and
GMOS on Gemini South (November 2003 and December 2004) resulted in 477
new spectra. These spectra were combined with 559 earlier
observations, velocities for which were re-computed to be consistent
with the new observations. 

We extract the online tables of Fornax objects and foreground Galactic
stars, and based on our velocity limit of 600\,\kms\ for the Fornax
cluster instead of the author's limit of 450\,\kms, we extract
velocities with Class A (secure velocity) classifications for 182
foreground stars and 554 Fornax GCs. The spatial extent of the
observations means that the GC population contains both objects
physically associated with NGC 1399 and objects within the
intra-cluster medium or associated with other massive host galaxies. 

\subsection{The final catalogue}

Table~\ref{tab:sources} lists the sources of redshifts contributing to
the compilation, and the number of redshifts contributed by each in
the final catalogue. 

\begin{table}
\centering
  \caption{Sources contributing to our spectroscopic compilation. The
    Source column gives the code for each work, as it appears in the full
  catalogue. }
\label{tab:sources}
\begin{tabular}{llrr} \hline
Source & Reference & Instrument & N(Spectra)  \\ \hline
1 & This work & AAOmega/2dF & 1553  \\
2 & Drinkwater 2dF & 2dF & 12071  \\
3 & Firth 2dF & AAOmega/2dF & 1608  \\
4 & Gregg 2dF & VLT/FLAMES & 25  \\
5 & Drinkwater FLAIR & FLAIR-II & 346 \\
6 & Firth FLAMES & FLAMES & 46  \\
7 & 2MRS & Multiple & 276  \\
8 & Hilker 1999 & du Pont & 29  \\
9 & Ferguson1989 & Multiple & 57  \\
10 & Bergond 2007 & FLAMES & 138 \\
11 & Mieske 2004 & du Pont & 102 \\
12 & Schuberth 2010 & FORS2/GMOS & 605 \\
Total &  & & 16856 \\  \hline
\end{tabular}
\end{table}

Table~\ref{tab:speccat} is the final list of 16856 objects with
spectroscopic redshifts. Only objects with `secure' redshifts,
as determined by each work, were included. For our spectroscopy and
those we reduced from \citet{Firth2008}, redshifts were considered
secure if assigned a quality operator (QOP) value of 3 or 4, which
indicates the redshift was derived either from a single, unambiguous
spectral feature, or several identifiable spectral features. Other
works have their own criteria for secure redshifts, and we follow
their guidelines for usage. Each object is assigned a
unique identifier. The four general populations of objects within the
table are Galactic stars, Fornax galaxies, Fornax GCs, and background
objects, including galaxies and quasars. In particular, we included
Galactic stars in the table to enable the reader to choose the
velocity limits of the Fornax cluster themselves.

Individual targets are often observed more than once. Duplicate
observations within studies, and duplicate observations spanning
multiple studies, have been removed from the catalogue. Duplicate
observations within a study are removed either by their object ID (preferred) or by
positional match, with a matching radius of 1\arcsec. A small matching
radius is appropriate as the observations are derived from the same input
imaging, so positional consistency between observations should be good.

A matching radius of 2\arcsec\ is used to identify duplicate observations across
multiple studies. This is a conservative radius, as for some of the 
larger, extended cluster members, the object centroid differs based on
the input photometry used by several arcseconds. Therefore, we caution
the reader that there may still be a few tens of objects in the
catalogue with duplicate entries. In the case of duplicate
observations, priority is given in the order listed in Table
~\ref{tab:sources}. 

\begin{figure}
\includegraphics[width=\columnwidth]{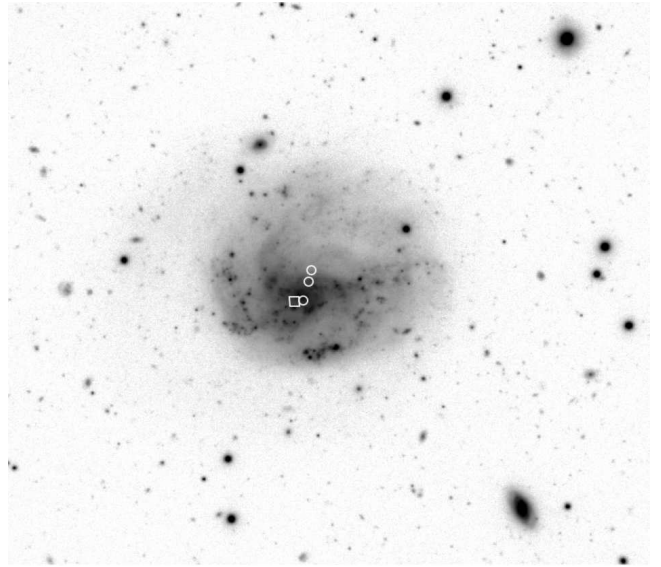}
\caption{Example of an extended galaxy with multiple spectra
  sufficiently far apart from each other that they are not identified
  as duplicate observations of the same galaxy. The white square
  indicates the position of the object from the
  \citet{Venhola2018} photometry catalogue, while the white
  circles are the coordinates of the three spectra. Each circle has a 
  2\arcsec\ radius, and the square is 4\arcsec\ on a side. The galaxy
  is NGC 1437A (FCC 285).}
\label{fig:ExampleGalaxy}
\end{figure}

Fig.~\ref{fig:ExampleGalaxy} illustrates one of these cases. There are
three spectra for the same object (marked as white circles) in the
spectroscopic catalogue, separated by more than the 2\arcsec\ radius
used to identify duplicate observations. The maximum separation
distance between these spectra is 13.5\arcsec, and the three spectra have
recession velocities $cz = $891, 891 and 895\kms. 

\begin{table}
  \centering
  \caption{The spectroscopic catalogue. The full catalogue is
    available online, a subset of 10 lines is given here for
    guidance. The ID is unique to this compilation, while the Source is as listed in
    Table~\ref{tab:sources}. The Type is as given in each
    work. Note that many of the GS have moved since the original
    observations were taken.}
  \label{tab:speccat}
  \resizebox{\columnwidth}{!}{
  \begin{tabular}{lcccccc} \hline
    ID & RA & Dec & $cz$ & $cz_{err}$ & Source & Type  \\
    & J2000 & J2000 & \kms & \kms & & \\ \hline
 F17029 &   53.7313 &  -35.0666 &    230 &    60 & 12 &  FGS \\
F17031 &   53.7402 &  -35.0564 &    195 &    45 & 12 &  FGS \\
F17033 &   53.7504 &  -35.1570 & 651000 & 45000 & 12 &  QSO \\
F17035 &   53.7628 &  -35.3482 &  49220 &   105 & 12 &  BGO \\
F17037 &   53.7637 &  -35.0094 &  94500 &   175 & 12 &  BGO \\
F17038 &   53.7661 &  -35.2338 &    180 &    25 & 12 &  FGS \\
F17042 &   53.7804 &  -34.9703 &    130 &    25 & 12 &  FGS \\
F17047 &   53.8263 &  -34.9727 &    160 &    35 & 12 &  FGS \\
F17051 &   53.8509 &  -35.2808 &    135 &    25 & 12 &  FGS \\
F17062 &   54.1800 &  -35.7606 &  30821 &    84 & 12 &  BGO \\ \hline
\end{tabular}}
\end{table}

\begin{figure}
\includegraphics[width=\columnwidth]{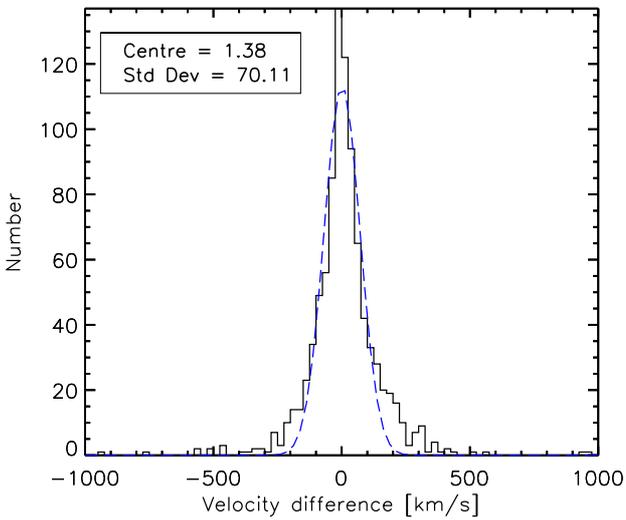}
\caption{Histogram of objects with more than one observation
  resulting in a secure recession velocity. A Gaussian function is fit
  to the distribution, centred very near zero and with standard
  deviation $\sim$70\,\kms. }
\label{fig:duplicates_vdiff}
\end{figure}

We can estimate the consistency of recession velocity determinations
between the various surveys by extracting objects that were observed more
than once. Selecting objects with secure measurements from more
than one survey, we find very good agreement in their velocities, as shown in
Fig.~\ref{fig:duplicates_vdiff}. The results do not change
substantially if the duplicate measures are divided into
three populations, i.e. Galactic stars, Fornax objects and background
objects. The uncertainties are not large enough for objects to
have their classification of foreground, Fornax or background to be in
question. Only 14 objects within the full catalogue have discrepant
redshifts $\Delta z > 0.1$, or $\Delta cz > 30\,000$\kms. Nearly all
of these are extragalactic objects mistakenly classified as Galactic
stars, five of which are broad-line quasi-stellar objects (QSOs). None are Fornax cluster
galaxies. Since we do not have the original spectra for these objects,
we cannot inspect them to understand the discrepancies.

While all coordinates given in Table~\ref{tab:speccat} are J2000 and
extracted from each catalogue, the many Galactic stars within the
catalogue may have moved substantially since the observations were
taken, by as much as several arcseconds. This should be taken into
account when using the table, in particular when crossmatching the
table with new photometry. 

No velocity uncertainties are output by the MARZ program, so the
velocities from this work, Firth 2dF and Gregg 2dF do not have velocity
uncertainties in the table. However, since all three sets of
observations use the same instrumentation, the uncertainties will be
similar for each. There are 166 objects from the new spectroscopy and
those of Firth 2dF that have two high confidence redshifts. The velocities of the
duplicate observations all agree to within $\pm$40\kms. 
From the Drinkwater 2dF spectra, the provided redshift uncertainties
are $\Delta z/(1+z)$=0.0002--0.0004, corresponding to velocity uncertainties
of 60--120\kms. These are all consistent with the estimated redshift
uncertainty of the MARZ program of $\sim$ 0.0003, determined by
comparison with an independent redshifting method \citep{Hinton2016}.

\section{The Fornax cluster and beyond}\label{sec:FCandbeyond}

\subsection{The Fornax volume}\label{subsec:fornaxvol}

\begin{figure}
\includegraphics[width=\columnwidth]{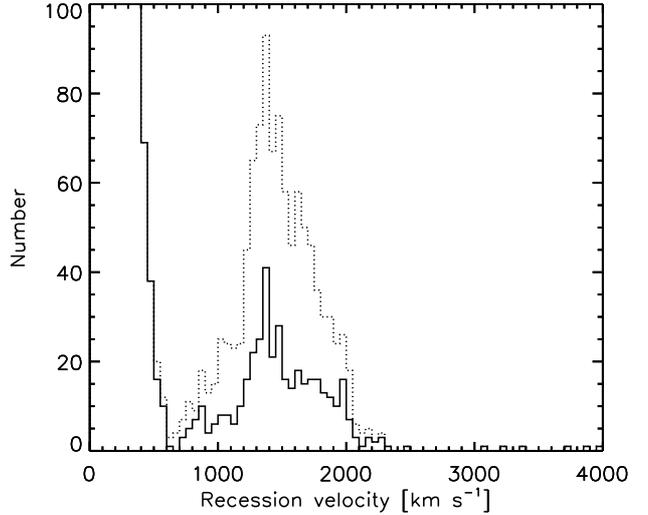}
\caption{Histogram of recession velocities within the Fornax cluster
  volume. The minimum at 600\,\kms\ marks the delimiting velocity
  between Galactic stars and the cluster. The Galactic stars peak
  extends to 2000 objects on this figure. The dotted line includes the
many hundreds of GCs with measured redshifts, whereas the solid line
includes only GSs and galaxies.}
\label{fig:Fornax_zhist}
\end{figure}

Fig.~\ref{fig:Fornax_zhist} shows the volume contained within
$0<cz<4000$\,\kms. We adopt a lower velocity limit of 600\,\kms\ to
separate Galactic stars from Fornax cluster galaxies, based on the
obvious minimum at this velocity, although lower values are occasionally
used in the literature. The volume behind the cluster is exceptionally
empty. We adopt 3000\,\kms\ as the upper limit for the Fornax cluster volume.

Of the 16856 unique objects with spectroscopic redshifts, 1039 belong within
the Fornax volume of $600<cz<3000$\,\kms, of which 651 are confirmed GCs,
leaving 388 cluster galaxies. 9483 are Galactic stars, reflecting the
fact that many works either specifically targeted point sources, or imposed no
morphological selection. While this approach results in a low yield of
extragalactic objects, this strategy revealed the population
of ultra-compact dwarf galaxies in the Fornax cluster.

\begin{figure}
\includegraphics[width=\columnwidth]{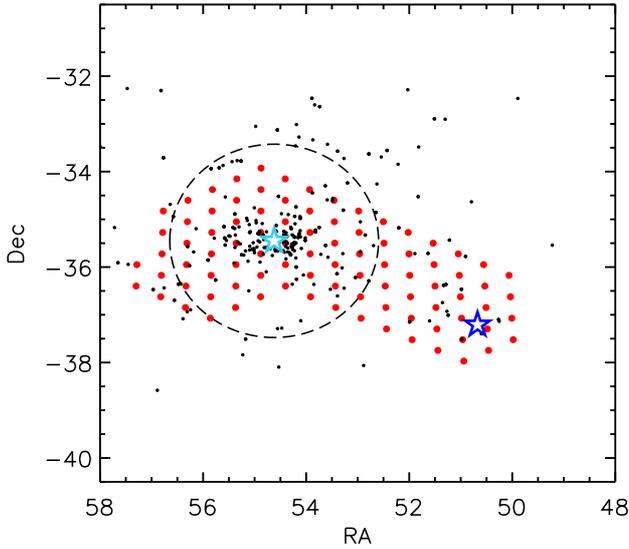}
\caption{Spatial distribution of the 388 confirmed Fornax cluster
  galaxies. The overdensity at RA=54.5, Dec=-35.5 is centred on
  NGC 1399, shown as a light blue star, and
  reflects a true overdensity of objects, as well as the preference of
  observing campaigns to focus on this inner cluster region. The dark
  blue star to the south-west marks the position of Fornax A. The
  black dashed circle outlines the approximate virial radius of the
  main cluster of 700 kpc. As in Fig.~\ref{fig:footprint}, the outline of the area
  to be covered by the MFS is shown with red points.}
\label{fig:fornax_space}
\end{figure}

Fig.~\ref{fig:fornax_space} shows how the Fornax cluster members are
distributed in space. GCs are not shown here, as they cluster heavily
around NGC 1399. There is a slight overdensity of objects in the
south-west extension, centred on Fornax A. 

\begin{figure}
\includegraphics[width=\columnwidth]{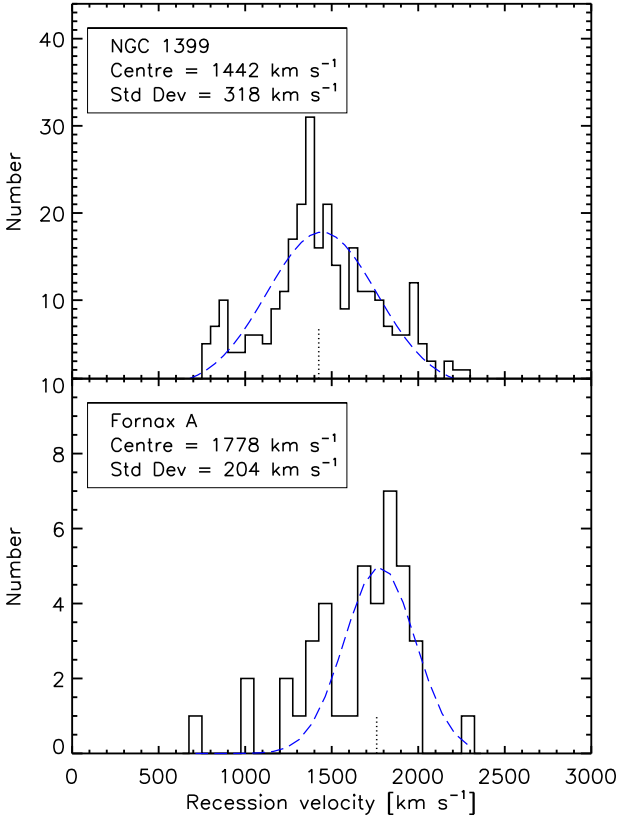}
\caption{(Top) Histogram of the recession velocities of the 294 objects that
  lie within 700 kpc of NGC 1399, excluding GCs (black solid
  histogram). The fit to the velocity distribution is centred at 1442\,\kms, with
  a width of 318\,\kms\ (blue dashed line). (Bottom) Histogram of the
  recession velocities of the 40 objects that lie within 700 kpc of
  Fornax A (black solid histogram), with the fit centred at a
  significantly higher velocity, 1778\,\kms. The short dotted marks at
  $cz=1425$\,\kms\ (top) and $cz=1760$\,\kms\ (bottom) indicate the
  recession velocities of NGC 1399 and Fornax A, respectively.} 
\label{fig:NGC1399_fornax_vhist}
\end{figure}

As we now have the most comprehensive census of spectroscopically
confirmed Fornax cluster members assembled to date, we can confidently
measure the velocity dispersion
of the Fornax cluster. Incorporating only objects with
$600<cz<3000$\kms\ and within the virial radius of 0.7 Mpc centred on
NGC 1399 as determined by \citet{Drinkwater2001b}, we 
find the cluster recession velocity is 1454\,\kms, offset from the
accepted recession velocity of NGC 1399 of 1425\,\kms\
(\citealt{Graham1998}). The width of the distribution is 286\,\kms.
Removing the GCs, which dominate the number counts and may not be
sampling the cluster potential, the cluster velocity is centred at
1442\,\kms, with a width of 318\,\kms. The distribution of velocities is
shown in the top panel of Fig.~\ref{fig:NGC1399_fornax_vhist}, along
with a Gaussian fit, as well as the recession velocity of NGC 1399 itself at
1425\,\kms. We note that more sophisticated methods of parametrising
the complex distribution of velocities could be used, but a simple
Gaussian is sufficient for our purposes here.

We can separate the objects in the top panel of
Fig.~\ref{fig:NGC1399_fornax_vhist} into dwarf galaxies and giant
galaxies, based on the photometry described in
Section~\ref{subsec:photometry}. For dwarf galaxies confirmed to have $M_r >
-18.5$, as used by \citet{Venhola2019}, and employing a distance modulus
of 31.51 as in 
\citealt{Blakeslee2009}, the centre and width of the recession velocity
distribution are 1435\,\kms\ and 303\,\kms, respectively. The results
are insensitive to the magnitude delimiting dwarfs from giants. 

This is much smaller than the 429\,\kms\ dispersion found in
\citet{Drinkwater2001b}, who use only the 55 dwarf galaxies within
their FLAIR-II observations. These 55 galaxies are a subset of the 168
dwarfs found in the current compilation, which more fully probe the
velocity distribution of the cluster. They are also uniformly
distributed across the volume of the cluster, whereas our 168 objects
are more concentrated around NGC 1399. 

The \citet{Drinkwater2001b}
study also uses $M_B>-16$ to define the dwarf regime, compared to
$M_r>-18.5$ (which is approximately $M_B>-18$ for colours typical of
dwarf galaxies). Relaxing this to $M_B>-18$ to be more comparable to
the magnitude limit used here, the resulting 84 galaxies in their
sample have lower velocity dispersion of 370\,\kms.

Giant galaxies are
selected to either have $M_r< -18.5$ or be from the 2MRS or Ferguson
1989 compilations, which both have very bright flux limits. There are
only a few tens of such bright galaxies in our survey area, so the
velocity distribution is not well-described by a Gaussian, but the
resulting fit is centred at 1453\,\kms, with a width of 329\,\kms, not
substantially different from the dwarfs or the general population.

Using the relation between the dynamical mass, $M_{dyn}$, of a cluster, and
the velocity dispersion, $\sigma_{v}$, as determined in \citet{Saro2013}:

\begin{equation}
M_{dyn} = \left( \frac{\sigma_{v}}{A \times h^{C}}\right) ^{B} 10^{15}
\, M_{\odot}
\label{eq:mass}
\end{equation}

\noindent with $A = 939$, $B = 2.91$ and $C = 0.33$, $h = 0.73$ and $\sigma_{v}
=318$\,\kms, we derive a dynamical mass of $5.8\times
10^{13}$\,\Msun. We caution that the relation in
equation~\ref{eq:mass} was calibrated for clusters with masses
$M_{dyn}\goa 10^{14}$\,\Msun, so we are extrapolating to lower masses.
\citet{Drinkwater2001b} find a velocity of the 
overdensity around NGC 1399 centred at 1478\,\kms\ with width
370\,\kms. However, they only had 92 spectroscopic cluster members, 
compared with our 945, or 294 excluding GCs. They employ a number of
methods of computing masses, with results spanning $5-9\times
10^{13}$\,\Msun, consistent with our estimate. 

We also have 40 galaxies confirmed to lie within 0.7 Mpc of Fornax A to
the south west. This number is small, but can provide an indication of the
subgroup velocity. Although the small numbers mean
the distribution is not well-fit by a Gaussian, the central velocity
of the Fornax A subgroup is found to be 1778\,\kms\ with 204\,\kms\
width, as seen in the bottom panel of
Fig.~\ref{fig:NGC1399_fornax_vhist}. This is in good agreement with
the recession velocity of Fornax A at 1760\,\kms. Again using
equation~\ref{eq:mass}, the velocity dispersion corresponds to a mass
of $1.6\times 10^{13}$\,\Msun. For the Fornax A subgroup, using only 16 galaxies,
\citet{Drinkwater2001b} find a recession velocity of 1583\,\kms, higher
than that of the main NGC 1399 overdensity, but lower than our value.

A two-sided Kolmogorov-Smirnov (KS) test on the recession velocities
for the structures centred on NGC 1399 and Fornax A, returns a
probability that the two distributions are drawn from the same
underlying sample of 8$\times$10$^{-5}$, so it is likely that the
Fornax A subgroup is kinematically distinct from the NGC 1399
overdensity, consistent with the conclusion drawn in \citet{Drinkwater2001b}.

Comparing the top and bottom panels of
Fig.~\ref{fig:NGC1399_fornax_vhist}, the NGC 1399 distribution appears
skewed toward high velocities, while the Fornax A distribution appears
skewed to low velocities. We have extracted the objects that lie more
than 1-$\sigma$ from the mean velocity (mean + 1-$\sigma$ for NGC 1399
and mean - 1-$\sigma$ for Fornax A) to explore the spatial
distribution of these objects. Although \citet{Venhola2019} find an
overdensity of dwarf galaxies in the region between NGC 1399 and
Fornax A, we find no evidence for similar behaviour of the general
galaxy population. 

\subsection{Clusters behind clusters}

Our spectroscopic compilation includes 6334 background galaxies,
extending to $z\sim 3$, with the majority of galaxies at $z<0.4$, and
the high redshift objects all broad-line quasars.

\begin{figure}
\includegraphics[width=\columnwidth]{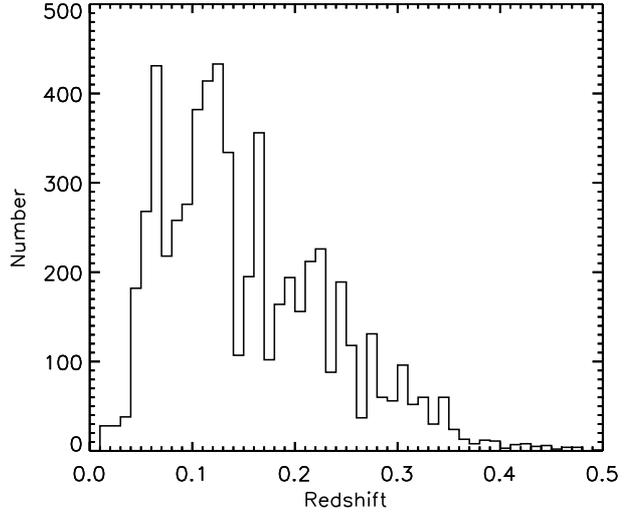}
\caption{Histogram of redshifts for objects in the volume behind the
  Fornax cluster. There are narrow overdensities at $z\sim 0.06$
  and $z\sim 0.17$ in the histogram, as well as the cluster at
  $z=0.11$, as noted by previous works. The 
  magnitude limits of the surveys prevent galaxies 
from being selected for observation beyond about $z\sim 0.5$, which is
well-matched to the MIGHTEE-\HI\ volume probed. There are $>$200
confirmed quasars extending to $z\sim 3$. }
\label{fig:Background_zhist}
\end{figure}

\begin{figure}
\includegraphics[width=\columnwidth]{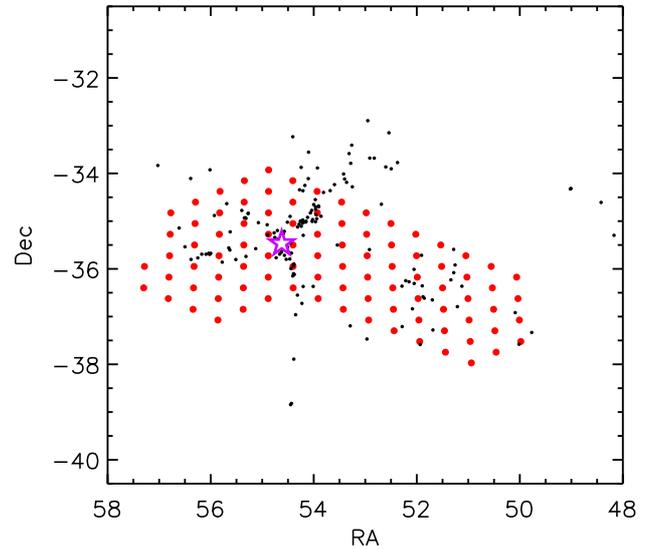}
\caption{Location of objects in the overdensity at 33\,000--34\,000\,\kms, or
  $z=0.11$. The BCG marked with a purple star lies at the intersection of 
  several filaments. The small number of objects in the extension toward
  Fornax A are probably physically unrelated.}
\label{fig:Background_cluster_space}
\end{figure}

From Fig.~\ref{fig:Background_zhist}, 
we confirm the overdensity at $z=0.11$ also reported in
\citet{Hilker1999b}. The peak velocity we find of 33\,528\,\kms\ agrees very
well with their value of 33\,580\,\kms, although we find a narrower
dispersion, only 144\,\kms, in contrast with their 316\,\kms.
\citet{Hilker1999b} identify the BCG of the cluster as CGF 1-1, only
1\farcm 1 south of NGC 1399, and lies at the intersection of at least
two filaments, as seen in Fig.~\ref{fig:Background_cluster_space},
indicated by the purple star. The filaments extend $\sim$4 degrees on
the sky in the North--South direction, which at the distance of this
overdensity, corresponds to a 
linear size of 29 Mpc. With our increased numbers of objects
(168 compared with 19 from \citealt{Hilker1999b}), we find that
the overdensity at this redshift is actually composed of several
smaller peaks. However, from Fig.~\ref{fig:Background_cluster_space},
several of the 168 objects are probably physically unrelated.

Also seen in Fig.~\ref{fig:Background_zhist}, there are two additional,
narrow overdensities at $z=0.05$ and $z=0.17$, which have velocity standard
deviations of 275 and 240\,\kms, respectively. From the spatial
distributions, all three overdensities appear to be filamentary in
structure.

\subsection{High redshift and radio sources}

There are 264 objects marked in the catalogue as QSOs, 
23 of which are within the MIGHTEE-\HI\ volume of
$z_{\mathrm{HI}}<0.58$, while 205 are at $z>1$.
The absence of colour selection within the Drinkwater 2dF
spectroscopic campaign was leveraged by \citet{Meyer2001} to construct
an unbiased QSO sample from the first epoch of
observations. Their comparison of the flux-limited sample with other
commonly used methods of identifying QSO candidates showed that
selecting objects with a UV-excess results in samples which are
69~per~cent and 50~per~cent complete at $0.3<z<2.2$ and $z>2.2$,
respectively. Multi-colour selection increased completeness to 90 and
80~per~cent at low and high redshifts, respectively.

To within a matching radius of
20\arcsec, only 12 of the 264 have matches in the NRAO VLA Sky Survey
radio catalogue (NVSS; \citealt{Condon1998}). A
large crossmatching radius is necessary due to the different location
of radio emission with respect to the optical host galaxy position for
double-lobed radio sources. The NVSS catalogue has a 
relatively bright peak flux limit of 2 mJy beam$^{-1}$, thus preferentially selects
only nearby star-forming galaxies and the most radio-bright
quasars. Table~\ref{tab:radio} lists the 86 objects at 
$z>0.1$ that are also detected in NVSS, 12 of which are QSOs. 

With the deep radio imaging from MeerKAT, we expect most, if not all, of
the QSOs in this field to be detected at radio wavelengths, even those
classically considered to be radio quiet \citep{White2015}. The MFS
will reach a root-mean-square (RMS) flux limit of $\sim 3 \mu$Jy,
nearly three orders of magnitude deeper than NVSS. These
objects can be used as background sources searching for either
associated or intervening absorption of \HI\ along the line of
sight. Having spectroscopic redshifts of the radio continuum sources
immediately enables us to determine whether any \HI\ absorption seen
in the radio spectrum is associated with the background source (the
absorption is at the same redshift as the continuum source) or
intervening along the line of sight to the background source (the
absorption is at a lower redshift than the continuum source).

\begin{table}
  \centering
  \caption{Objects at $z>0.1$ also detected in NVSS. The full table is available
  in the online version.}
  \label{tab:radio}
  \begin{tabular}{lcccc} \hline
    ID & RA & Dec & Redshift & Peak flux \\
        & J2000 & J2000 &              & [mJy/beam] \\ \hline
F00057 &     50.2941 &    -36.2826 & 0.294 &   2.029 \\
F00107 &     50.6439 &    -37.3849 & 0.168 &   4.454 \\
F00112 &     50.6744 &    -36.8857 & 0.356 &   5.540 \\
F00208 &     51.1021 &    -36.6014 & 0.361 &   3.149 \\
F00221 &     51.1596 &    -37.3265 & 0.169 &  10.398 \\
F00249 &     51.2431 &    -36.7258 & 0.433 &  58.723 \\
F00322 &     51.4887 &    -36.4247 & 0.272 &   6.727 \\
F00498 &     52.2118 &    -37.7718 & 0.114 &   2.113 \\
F00535 &     52.3344 &    -36.9957 & 0.103 &   2.135 \\
F00554 &     52.3832 &    -36.7790 & 0.326 &   6.450 \\ \hline

\end{tabular}
\end{table}

\subsection{Photometry}\label{subsec:photometry}

In order to understand the magnitude range of observed objects, the
spectroscopic catalogue is matched to the output of Source 
Extractor over the Fornax region from the FDS, which served as the
initial object detection list for the work in \citet{Venhola2018}, and
thus includes both resolved and unresolved source. Of 16856 objects,
15254 have a counterpart in the optical catalogue to a crossmatching
radius of 1\farcs 5. The objects without 
optical counterparts are primarily composed of five categories. The
first is objects with a spectrum that lie outside the imaging footprint of the
FDS. The second is Galactic stars. As the time difference between the
spectroscopic observations and the new FDS imaging is as much as
$\sim$20 years in some cases, the stars have moved on the sky by
several arcseconds.

The third group of objects often without photometry are the CSS objects. As
these primarily lie within the halos of more massive galaxies,
specialised photometry processing is required to isolate these
objects. The next group are the massive galaxies themselves, which
require photometry processing tuned toward spatially extended
objects. There are a few tens of such large galaxies in our sample.
The final group is background galaxies, with an increased
incompleteness behind NGC 1399, where the density of foreground
galaxies is high, and blending of the photometry is an issue.

\begin{figure}
\includegraphics[width=\columnwidth]{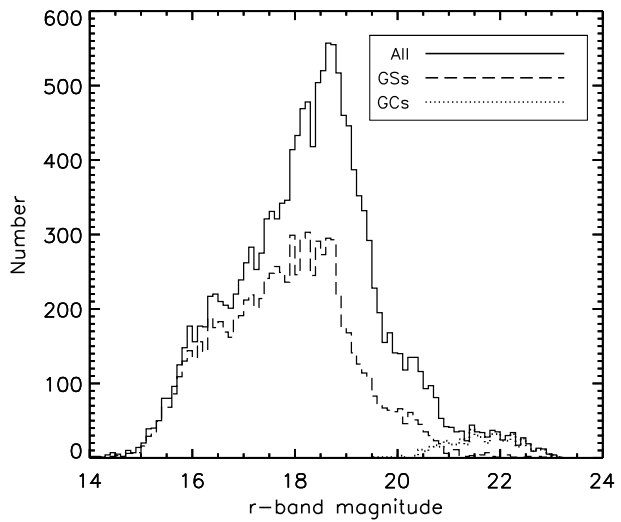}
\caption{Histogram of $r$-band magnitudes of objects in the
  sample. The solid histogram is the full sample, the dashed histogram
shows the magnitudes of confirmed Galactic stars. The dotted histogram
at the faintest magnitudes are GCs.}
\label{fig:rmag_hist}
\end{figure}

Fig.~\ref{fig:rmag_hist} shows the distribution of $r$-band magnitudes
from the FDS for objects with spectroscopy. The bulk of the objects are Galactic
stars. The faint end of the distribution is dominated by GCs, which
were targeted specifically, in particular by \citet{Schuberth2010},
who were able to extend to faint magnitudes due to the large
collecting area of the VLT used for the observations.

\subsection{Completeness}

We provide here an estimate of the completeness of the spectroscopy
over the FDS 26 deg$^2$ area. As the primary data product of this work
is the spectroscopic catalogue, we only perform a basic completeness
calculation for interest.

Computing the overall spectroscopic completeness is complicated by a
number of competing 
factors. The most severe is the high density of targeted observations focusing
exclusively around NGC 1399, resulting in very high spectroscopic completeness in
that area, with lower completeness elsewhere. The high density of
foreground objects around NGC 1399 also results in the difficult
crowded-field photometry, affecting the completeness of background
galaxies. Another main issue is the missing GSs and Fornax CSS systems
from the crossmatch between the spectroscopy and photometry.

We eliminate the missing point source complication by employing the photometric
catalogues from \citet{Venhola2018} which have a restriction for the
minimum size of included objects, with semi-major axis
$\ge$2\arcsec. This removes the GSs and CSS objects, but is small enough
to include objects similar in size to Local Group dwarf spheroidals at
the distance of the Fornax cluster. Additional objects removed by the
size restriction are background QSOs and compact background
galaxies. However, the catalogue is a fair representation of the
extragalactic population, particularly at the distance of the Fornax
cluster, the primary focus of many studies.

The 2\arcsec\ photometry catalogue is crossmatched to the
spectroscopic catalogue, with magnitude-based crossmatching radii. At
bright magnitudes, $R<14$, the galaxies are large, and the centroid of
the objects is dependent on the depth of the photometry and the
processing employed. Fig.~\ref{fig:ExampleGalaxy} illustrates the
issue of using different 
photometry as the input for spectroscopic observations. The white
square indicates the coordinates of the object as listed in our input
catalogues, while the white circles show where spectroscopic
observations were centred. The nearest spectrum is 4\arcsec\ distant
from the photometric coordinates, and is thus missed by a small positional
crossmatch between photometry and spectroscopy. For these objects, we
use a crossmatching radius of 15\arcsec\ between the galaxy position
and the corresponding spectroscopy.

\begin{figure}
\includegraphics[width=\columnwidth]{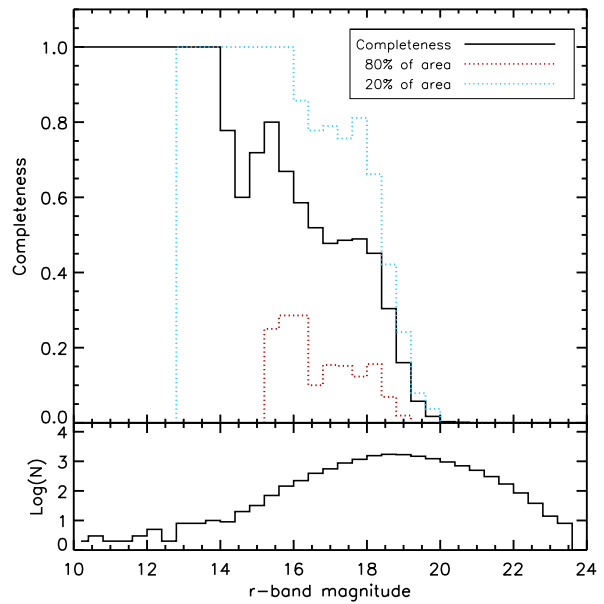}
\caption{(Top) Completeness of the spectroscopy as a function of FDS $r$-band
  magnitude. The solid black line indicates the average completeness
  over the full 26deg$^2$ FDS area, which varies strongly as a
  function of location. The red dotted line indicates the completeness
  achieved by 80~per~cent of the survey area, and the blue dotted line
  shows the completeness reached by 20~per~cent of the area. 
  (Bottom) Number of objects in the photometric catalogue
  as a function of magnitude.}
\label{fig:completeness}
\end{figure}

At fainter magnitudes, we use a crossmatching radius of
1\farcs 5. The spectroscopic completeness as a function of $r$-band
magnitude is shown in Fig.~\ref{fig:completeness}. We achieve
100~per~cent completeness for the bright galaxies. There is a small
bump at $r\sim 18$, which reflects the approximate magnitude limit of
many of the contributing spectroscopic surveys. The variation of
completeness depending on location within the cluster is also
indicated. The red dotted line shows the completeness reached by
80~per~cent of the survey area, while the blue dotted line is the
completeness reached by 20~per~cent of the area. 

Difficulty of obtaining spectroscopic redshifts for low surface
brightness galaxies is also a source of incompleteness. The behaviour
of central surface brightness as a function of magnitude for the input
catalogue for the Drinkwater 2dF spectroscopy is illustrated in Figure
2 of \citet{Drinkwater2000a}. The VST imaging serving as input
photometry for the spectroscopy in this work shows similar behaviour,
but extends to fainter magnitudes. As stated in \citet{Deady2002},
two-hour integrations with the original 2dF instrument reached
80~per~cent completeness to surface brightness of 23.2 mag
arcsec$^{-2}$. There are very few objects within the magnitude limits
probed here that have surface brightnesses fainter than about 23 mag
arcsec$^{-2}$, so while they are an interesting subpopulation of
objects, they are only a small source of overall incompleteness.

\section{Discussion and Summary}\label{sec:discussion}

While this is the most comprehensive compilation of spectroscopy in
the Fornax cluster region to date, we make no strong claims on the
completeness with respect to any given population of object. Each of
the spectroscopic campaigns imposed different selection criteria,
involving imaging from different sources, morphological
classification, colours and location within the cluster. The region
surrounding NGC 1399 understandably has the most spectroscopic
observations. The
observations of \citet{Drinkwater2000a} are close to unbiased, as the
only selection criterion was a magnitude limit. However, the large
fraction of confirmed Galactic stars (7969 of 12071 spectra) means this
approach is inefficient and requires large amounts of telescope
time. A magnitude limit also introduces a bias against objects with
low surface brightness, but the lack of morphology cut resulted in the
discovery of the UCD population of galaxies.

Nonetheless, the spectroscopic compilation provided here will be of great value for
studies targeting the cluster itself, and also the background
volume. The inclusion of the many thousands of confirmed Galactic
stars and GCs, in addition to the new high-quality imaging from the FDS, will
assist in future observing campaigns aimed at morphologically compact
objects, as target selection can incorporate knowledge of the colour
space occupied by the different populations.

With our vastly increased numbers, we can constrain the recession
velocity of the cluster centred on NGC 1399 to be 1454\,\kms, or
1442\,\kms\ if the GC population is excluded. This is distinct from
the substructure centred on Fornax A, which is found to have a higher
velocity of 1778\,\kms. We also confirm a background structure at $z=0.11$
that was hinted at in previous observations, along with at least two
other overdensities, which should be interesting environments to
explore with the upcoming radio spectral line and continuum surveys
with MeerKAT.

With the onset of the MFS and MIGHTEE surveys, which are
spectroscopic observations, we expect
additional galaxies to have spectroscopic redshifts derived from their
\HI\ profiles. This will be particularly true for low surface brightness,
but \HI-rich, objects, for which an optical spectrum is difficult to
obtain, but within reach of the excellent sensitivity of
MeerKAT. These objects can be added to the census of redshifts as they
become available. 

Due to its dynamic nature, ongoing growth, and relative proximity, the
Fornax cluster has been a rich environment for galaxy evolution
studies for several decades. With the new observing facilities in the
Southern hemisphere, interest has been renewed.


\section*{Acknowledgements}

We thank the anonymous referee for a rapid response and helpful
comments which improved this paper. 

We are grateful to Michael Drinkwater for providing his catalogue of
spectroscopic classifications. NM would like to thank Samuel Hinton
for assistance with the MARZ tool. We would like to thank J. Mohr,
J. Allison, and T. Westmeier for useful discussions.

NM acknowledges support from the BMBF D-MeerKAT award 05A2017.
This project has received funding from the European Research Council
(ERC) under the European Union's Horizon 2020 research and innovation
programme (grant agreement no. 679627; project name FORNAX). 
AV acknowledges the Scholarship Fund of the University of Oulu for
the financial support. E.I. acknowledges financial support from the
VST project (P.I. P. Schipani) and from the European Union Horizon
2020 research and innovation programme under the Marie Skodowska-Curie
grant agreement n. 721463 to the SUNDIAL ITN network. SIL is supported
by the National Research Foundation (NRF) of South Africa.

This research has made use of NASA's Astrophysics Data System
Bibliographic Services. This research has made use of the VizieR
catalogue access tool, CDS, Strasbourg, France (DOI: 10.26093/cds/vizier). 
Based in part on data acquired through the Australian Astronomical
Observatory, under program A/2017B/006.
This paper includes data that has been provided by AAO Data Central
(datacentral.org.au).









\bsp	
\label{lastpage}
\end{document}